Magnetic-field-insensitive coherent-population-trapping resonances excited by bichromatic linearly polarized fields on the $D_1$ line of $^{133}$Cs

Kenta Matsumoto[1,2], Sota Kagami[1,2], and Akihiro Kirihara[1,2]
[1]System Platform Research Laboratories, NEC Corporation, 1753 Shimonumabe, Nakahara-ku, Kawasaki, Kanagawa 211-0011, Japan
[2]NEC-AIST Quantum Technology Cooperative Research Laboratory, National Institute of Advanced Industrial Science and Technology (AIST), 1-1-1 Umezono, Tsukuba, Ibaraki 305-8568, Japan

Shinya Yanagimachi[3]
[3]National Institute of Advanced Industrial Science and Technology (AIST), 1-1-1 Umezono, Tsukuba, Ibaraki 305-8563, Japan

Takeshi Ikegami[4] and Atsuo Morinaga[4]
[4]MNOIC, Micromachine Center, AIST Tsukuba East 4G, 1-2-1 Namiki, Tsukuba, Ibaraki 305-8564, Japan

(Abstract)

We have experimentally demonstrated that magnetic-field-insensitive coherent-population-trapping (CPT) resonances are generated between the ground hyperfine levels on the $D_1$ line of $^{133}$Cs using a two-photon $\Lambda$ scheme excited by lin ∥ lin polarizations. The frequency shift of the CPT resonance is 0.04 Hz for the deviation of 1 μT at a "magic" magnetic field of 139 μT and is 50 times smaller than that of the conventional clock transition at a typical bias magnetic field for the clock operation. The amplitude of the CPT spectrum excited by lin ∥ lin polarizations is enhanced as the excitation intensity increases and is well explained by the three-level model without trap levels. Thus, the CPT resonance on the $D_1$ line of $^{133}$Cs atom excited by a simple lin ∥ lin scheme will be one of the best candidates for frequency reference of miniature atomic clocks.

## I. INTRODUCTION

Coherent-population-trapping (CPT) resonance is a quantum interference phenomenon observed using a two-photon $\Lambda$-type scheme in which the interference signal exhibits a high Q-factor at the microwave transition frequency between quantum states, such as the hyperfine transition of an alkali atom [1]. Because of its availability as a simple optical setup, CPT is a key technology for miniature atomic clock devices [2]. To satisfy evolving demand for high-speed communications and global positioning, CPT-based miniature atomic clocks need high performances in terms of long-term stability [3,4], low-power consumption [5], and robustness to variations in their circumstances [6]; however, such performances tend to degrade when miniaturizing devices [7].

The CPT resonance frequency in an alkali atom is equal to the transition frequency between two magnetic sublevels selected from each of the low and high energy levels of the two hyperfine ground states. Here, their magnetic quantum numbers are denoted by $m_g$ for the lower energy level and $m_e$ for the higher energy level. The two-photon $\Lambda$-type scheme based on a $(m_g, m_e) = (0, 0)$ transition is commonly used as the frequency reference for the CPT clock because its resonance frequency is insensitive to the magnetic field in first-order approximation. The CPT resonance for the (0, 0) transition can be simply observed by circular-polarized two-photon $\sigma^+ - \sigma^+$ or $\sigma^- - \sigma^-$ excitation. In this case, however, some atoms are optically pumped to "trapped" levels not involved in the CPT resonance, resulting in a lower signal contrast [8]. Several techniques have been developed to enhance the signal contrast of the CPT resonance for the (0, 0) transition, such as the push-pull optical pumping scheme that uses two alternating circularly polarized fields with an optical path difference of one half of the clock wavelength [9], the $\sigma^+ - \sigma^-$ scheme using counter-propagating fields with opposite circular polarization [10], and the lin $\perp$ lin configuration using mutually orthogonal linearly polarized lights [11]. Though these methods can achieve high-contrast signals, the complexity of the optical setup for constructing optical pumping fields is a drawback. Additionally, these methods induce the (-1, 1) and (1, -1) CPT resonances, together with the (0, 0) resonance, so that the pileup of spectra may reduce its Q-value and shift the center frequency [12]. Further, the resonance frequency of the (0, 0) transition is shifted under the quantization magnetic field due to the second-order Zeeman effect [13].

In addition to the (0, 0) transition, we find other transitions between the ground hyperfine levels of alkali atoms whose first-order magnetic-field sensitivities can be negligible. The (-1, 1) transition is insensitive to the variation of the magnetic field at a specific "magic" magnetic field, where the two states consisting the transition undergo the same Zeeman shift. Up to the present, magnetic-field-insensitive transitions from $m_g = -1$ to $m_e = 1$ in some alkali atoms have been generated using the two-photon microwave-radiofrequency transition and have been successfully used as an ideal qubit [14], Ramsey spectroscopy [15], and atom interferometer [16].

In 2005, Taichenachev *et al.* reported the high-contrast CPT resonance of the (-1, 1) transition on the $D_1$ line of $^{87}$Rb atoms achieved by the two-photon $\Lambda$-type scheme with linearly polarized bichromatic fields (lin ∥ lin), which is suited for miniature atomic clocks [17,18]. The performance of the CPT resonance on the $D_1$ line of $^{87}$Rb atoms using several excitation schemes was investigated and summarized by Warren *et al.* [19]. Though Taichenachev *et al.* pointed out that only alkali atoms with a nuclear spin of $I = 3/2$ have the possibility of forming high-contrast magnetic-insensitive CPT resonance [17], in 2009, Watabe *et al.* observed high-contrast single CPT resonance in $^{133}$Cs with $I = 7/2$ using the lin ∥ lin scheme [20]. In the latter work, they did not identify the properties of the CPT transition and did not examine how the magnetic field affected its resonance frequency. Also in 2009, a single CPT resonance of Cs observed using the lin ∥ lin scheme was utilized practically in an atomic clock by Boudot *et al.* [21]. Previously, the (-1, 1) and (1, -1) transitions were observed as the quadrupole-coupled doublet spectra in Cs atoms [22]. However, up to date, the spectral structure of the CPT resonances of Cs for the lin ∥ lin scheme has not been clarified in detail.

In this paper, we show that the two CPT resonances for (-1, 1) and (1, -1) transitions of $^{133}$Cs atoms are excited separately in frequency by the lin ∥ lin scheme, and the CPT resonance for the (-1, 1) transition is insensitive to variations of magnetic fields in the vicinity of the magic magnetic field. We also show that the observed CPT resonance signal for (-1, 1) and (1, -1) transitions excited by the lin ∥ lin scheme increases with the excitation intensity, while the signal for the (0, 0) transition excited by the $\sigma - \sigma$ scheme saturates. Thus, we propose that the CPT resonance for the (-1, 1) transition excited by the lin ∥ lin scheme is a good candidate for frequency reference of a miniature atomic clock because of its high stability against magnetic fluctuations, the large signal amplitude at high light intensity, and the compactness of the light source.

## II. CPT RESONANCE of Cs

To discuss details of the CPT resonances on the $D_1$ line of Cs atoms under a magnetic field, the energy levels of the $6^2S_{1/2}$ ground and the $6^2P_{1/2}$ excited hyperfine states of $^{133}$Cs with the nuclear spin of $I = 7/2$ are shown in Fig. 1. The total atomic angular momentum of the ground $F$ and the excited $F'$ takes the value of 3 or 4. Here, we denote the two ground states $|6^2S_{1/2}\ F=3\rangle$ and $|6^2S_{1/2}\ F=4\rangle$ as $|g\rangle$ and $|e\rangle$, respectively, and the excited state $|6^2P_{1/2}\ F'=3\text{ or }4\rangle$ as $|i\rangle$. Under a magnetic field, the degeneracy of $|g\rangle$ ($|e\rangle$) is solved in accordance with the value of the magnetic quantum number $m_g$ ($m_e$). The order of energy splitting are reversed for $|g\rangle$ and $|e\rangle$ states because their $g$-factors have opposite signs for alkali atoms. In the following, we show how each magnetic sublevel of the ground states varies in relation to the magnetic field.

We denote energies of the magnetic sublevels $|g, m_g\rangle$ and $|e, m_e\rangle$ under a weak magnetic field $B$ as $hf_{m_g}(B)$ and $hf_{m_e}(B)$, respectively, where $h$ is the Planck constant. The transition frequency $\Delta f_{m_g,m_e}(B) = f_{m_e}(B) - f_{m_g}(B)$ between the two ground states, $m_g$ and $m_e$, can be described using the second-order approximation of the Breit-Rabi formula [12], as follows:

$$f_{m_g}(B) - f_{m_g}(0) = -\frac{g_J - 9g_I}{8}\mu_B m_g B - \frac{(g_J - g_I)^2}{4f_{\text{HFS}}}\left(1 - \frac{m_g^2}{16}\right)\mu_B^2 B^2 \quad (1)$$

$$f_{m_e}(B) - f_{m_e}(0) = \frac{g_J + 7g_I}{8}\mu_B m_e B + \frac{(g_J - g_I)^2}{4f_{\text{HFS}}}\left(1 - \frac{m_e^2}{16}\right)\mu_B^2 B^2 \quad (2)$$

where $\Delta f_{0,0}(0) \equiv f_{\text{HFS}} = 9.192631770$ GHz, and $g_I$, $g_J$, and $\mu_B$ represent nuclear $g$-factor, fine structure Landé $g$-factor, and Bohr magneton, respectively. The compensation for the higher order can be ignored for the magnetic field of $|B| < 500$ μT because it is less than 0.002%. $\Delta f_{0,0}(B)$, which is the frequency of CPT resonance for the (0, 0) transition, is

$$\Delta f_{0,0}(B) = f_{\text{HFS}} + \frac{(g_J - g_I)^2}{2f_{\text{HFS}}}\mu_B^2 B^2. \quad (3)$$

Because $|g_I|$ is about 5,000 times smaller than $|g_J|$, when $m_e = -m_g$, the first-order term in $\Delta f_{m_g,-m_g}$ is nearly cancelled. The sign of the first-order magnetic field coefficient of $\Delta f_{m_g,m_e}$ depends on the values of $m_g$ and $m_e$, although the second-order term is always positive. For the CPT resonance for the (-1, 1) transition, the first-order term is negative for $B > 0$. This means the value of $\Delta f_{m_g, m_e}$ reaches its minimum at a specific magnetic field strength, which we call the "magic" magnetic field. Using recent values of the physical constants [23], the frequency of the CPT resonance for the (-1, 1) transition is given by

$$\Delta f_{-1,1}(B) / \text{kHz} = f_{\text{HFS}} / \text{kHz} + 40.073758(8) \times 10^{-6} \left( B / \mu\text{T} - 139.3046(2) \right)^2 - 0.777662(3) \quad (4)$$

At $B = 139$ μT, the resonance frequency for the (-1, 1) transition of $^{133}$Cs is insensitive against magnetic field fluctuations. The existence of the magic magnetic field for the (-1, 1) transition is general for other alkali atom species such as $^{87}$Rb [15] or $^{23}$Na [24].

Here, we consider two-photon transitions between the two hyperfine levels of the ground state and one of the excited hyperfine levels, forming the $\Lambda$ scheme on the $D_1$ line of $^{133}$Cs. Cs atoms are excited by bichromatic lights that consist of the two optical angular frequencies $\omega_1$ and $\omega_2$, near resonant to $D_1$ transitions from $|g, m_g\rangle$ and $|e, m_e\rangle$, respectively. We assume two linearly polarized lights co-propagating along the z-axis, which is the direction of the longitudinal magnetic field. The linearly polarized electric field $\vec{E}_1(t; \omega_1)$ is parallel to the x-axis, and $\vec{E}_2(t; \omega_2)$ forms an angle $\theta$ with the x-axis. We use the standard vector basis defined as $\hat{e}_\pm = \mp(\hat{e}_x \pm i\hat{e}_y)/\sqrt{2}$ [25]. Then, the linearly polarized fields are described using the $\sigma^+$ and $\sigma^-$ circular polarized components, as

$$\vec{E}_1(t) = \frac{\varepsilon_1 \left( e^{i\omega_1 t} + e^{-i\omega_1 t} \right)}{2} \hat{e}_x = \frac{\varepsilon_1 \left( e^{i\omega_1 t} + e^{-i\omega_1 t} \right)}{2\sqrt{2}} \left( -\hat{e}_+ + \hat{e}_- \right) \quad (5)$$

$$\vec{E}_2(t) = \frac{\varepsilon_2 \left( e^{i\omega_2 t} + e^{-i\omega_2 t} \right)}{2} \left( \hat{e}_x \cos\theta + \hat{e}_y \sin\theta \right) = \frac{\varepsilon_2 \left( e^{i\omega_2 t} + e^{-i\omega_2 t} \right)}{2\sqrt{2}} \left( -e^{i\theta} \hat{e}_+ + e^{-i\theta} \hat{e}_- \right), \quad (6)$$

where $\varepsilon_1$ and $\varepsilon_2$ represent the amplitudes of electric fields, and the phase difference between the two fields is assumed to be zero for the sake of simplicity. The configurations of the excitation light, called lin ∥ lin and lin ⊥ lin excitation, are represented when $\theta = 0$ and $\pi/2$, respectively.

The Rabi frequencies $\Omega_{\omega_1}$ of photon $\omega_1$ for the transition from $|g, m_g\rangle$ to $|F', m_{F'}\rangle$ and $\Omega_{\omega_2}$ of photon $\omega_2$ for that from $|e, m_e\rangle$ to $|F', m_{F'}\rangle$ are written as follows:

$$\Omega_{\omega_1} = \frac{d_{3F'}\varepsilon_1}{2\sqrt{2}\hbar} \sum_{q_g=\pm 1}(-q_g)\langle F',m_{F'}|3,1,m_g,q_g\rangle \qquad (7)$$

$$\Omega_{\omega_2} = \frac{d_{4F'}\varepsilon_2}{2\sqrt{2}\hbar} \sum_{q_e=\pm 1}(-q_e)e^{-iq_e\theta}\langle F',m_{F'}|4,1,m_e,q_e\rangle. \qquad (8)$$

$d_{FF'}$ is the reduced matrix elements for the dipole moment operator between $F$ and $F'$, which are independent of $m_g$, $m_e$, and $m_{F'}$. The Clebsch-Gordan coefficient $\langle F',m_{F'}|F,1,m_F,q\rangle$ vanishes unless the magnetic quantum number of sublevels satisfy $m_{F'}=m_F+q$, where $q=+1$ for the $\sigma^+$ circular polarized light and $q=-1$ for the $\sigma^-$ circular polarized light propagating along the quantization axis. Therefore, the two-photon resonance between the two ground states $|g,m_g\rangle$ and $|e,m_e\rangle$ excited by bichromatic linearly polarized light is achieved with a relation of $|m_e-m_g|=0$ or 2, when $(\omega_1-\omega_2)/2\pi = \Delta f_{m_g,m_e}(B)$. The allowed combinations of $(m_g, m_e)$ are $(m, m)$, $(m-1, m+1)$, and $(m+1, m-1)$. The typical four $\Lambda$ schemes with $m=0$ are shown in Fig. 1, whose resonance frequencies are insensitive to the strength of the magnetic field.

The CPT resonance in a $\Lambda$ scheme occurs when interference appears in the excitation process, coherence is created in the two ground states, and the ensemble is placed in a non-absorbing state called a dark state. The dark state $|dark\rangle$ is described by a normalized linear superposition of the hyperfine states $|g,m_g\rangle$ and $|e,m_e\rangle$, as follows:

$$|dark\rangle \equiv \frac{a}{\sqrt{a^2+b^2}}\left(|g,m_g\rangle + \frac{b}{a}|e,m_e\rangle\right) = \frac{\Omega_{\omega_2}}{\sqrt{|\Omega_{\omega_1}|^2+|\Omega_{\omega_2}|^2}}\left(|g,m_g\rangle - \frac{\Omega_{\omega_1}}{\Omega_{\omega_2}}|e,m_e\rangle\right). \qquad (9)$$

Table I shows allowed two-photon $\Lambda$-scheme transitions for the CPT resonances on the $D_1$ line of $^{133}$Cs with the polarization scheme for the excitation and the magnetic quantum numbers, together with the dark state condition $b/a$ in the case of $m=0$ and $\varepsilon_1=\varepsilon_2$. The CPT resonances are classified into three independent transitions of $(m, m)$, $(m-1, m+1)$, and $(m+1, m-1)$. In the $(m, m)$ transition, there are four $\Lambda$ schemes in terms of the polarization scheme and the excited state. In the $(m-1, m+1)$ and $(m+1, m-1)$ transitions, there are each two $\Lambda$ schemes in terms of the excited states. The dark state of $(m_g, m_e)$ exists and the CPT resonance is observable only when all allowed $\Lambda$-scheme transitions under a given experimental condition have an identical value of $b/a$.

For example, in the (0, 0) transition via the excited state of $F'=3$, there are two different $\Lambda$ schemes excited by $\sigma^+-\sigma^+$ polarization and $\sigma^--\sigma^-$ polarization. A common dark state for the double $\Lambda$ schemes exist, when $e^{i\theta} = -e^{-i\theta}$. Thus, the CPT resonance signal for the (0, 0) transition excited by bichromatic linear polarizations is enhanced in the lin $\perp$ lin ($\theta = \pi/2$) configuration, while it disappears in the lin $\parallel$ lin ($\theta = 0$) configuration. On the other side, in the (0, 0) transition excited by $\sigma^+ - \sigma^+$ polarization, there are double $\Lambda$ schemes via the excited states of $F' = 3$ and $F' = 4$ in a vapor cell where the optical transitions are broadened by the Doppler effect and buffer-gas collisions [26]. The (0, 0) CPT resonance signal is enhanced, if values of two $\theta$ are the same. For both of $(m-1, m+1)$ and $(m+1, m-1)$ transitions via one of the excited states, the CPT resonances are observable at whatever value of $\theta$ in the bichromatic linear polarizations. It should be noted that the dark state of $(m-1, m+1)$ or $(m+1, m-1)$ built with the $F'' = 3$ excited state is collapsed by coupling with the other hyperfine excited level of $F'' = 4$ where the optical transitions are broadened.

Taichenachev *et al.* experimentally showed a pair of large spectra of (-1, +1) and (+1, -1) resonances on the $D_1$ line of $^{87}$Rb in lin $\parallel$ lin excitation and proposed that the excitation scheme is suited for a miniature atomic clock [17]. They noted that the $(-F, -F)$ and $(F, F)$ resonances can also have large contrast under excitation to a lower excited hyperfine state. Additionally, they predicted that the dark state for the (-1, 1) and (1, -1) transitions are not induced for atoms with any $F'$ state larger than $F' = 2$, including Cs atoms.

III. EXPERIMENTAL SETUP

Figure 2 shows the experimental setup used for detection of CPT resonances with a Cs-vapor cell. This setup can provide the two-photon excitation scheme whose polarizations are in $\sigma - \sigma$ or lin $\parallel$ lin configurations with or without a quarter-wave plate.

The light source is a distributed Bragg reflector (DBR) laser, which emits a single frequency light whose wavelength is tuned to the vicinity of the $D_1$ line of $^{133}$Cs at a wavelength of 894.59 nm. The laser light is injected into polarization-maintaining optical fiber and passes through an electronic variable optical attenuator (VOA) to control the output laser power. The output from the single frequency laser is modulated by a LiNbO$_3$ electro-optical modulator (EOM) driven with a modulation signal from a signal generator. The modulation frequency $f_m$ is set to around 4.596 GHz so that the frequency difference between +1st-order and -1st-order sidebands is nearby $f_{HFS}$. With the modulation index of 1.0, 20% of the total laser intensity is assigned to the +1st-order or -1st-order sideband, respectively. The frequency-modulated light is expanded to be a 7-mm-diameter beam and passes through a Cs-vapor cell. The cross-sectional shape and intensity of the beam are measured by a beam profiler and power meter installed at the location of the cell. On the beam path before the cell, a half wave plate is set to obtain the linearly polarized light. For $\sigma-\sigma$ polarization, a quarter-wave plate is additionally set. The total light intensity of the +1st and -1st order sidebands at the cell can be varied by VOA up to $5\ \mu W/mm^2$.

The Cs-vapor cell, 20 mm in diameter and 25 mm in length, is prepared at room temperature, which contains $^{133}Cs$ and 0.13 kPa (1 Torr) of $N_2$ buffer gas. The cell is placed inside a solenoid coil that can generate a longitudinal magnetic field parallel or antiparallel to the beam propagation direction up to $500\ \mu T$. The solenoid coil is covered by a magnetic shield to reduce stray magnetic fields from the circumstances. The strength of the magnetic field was measured using a fluxgate magnetometer within an accuracy of $0.3\ \mu T$.

The transmitted beam is introduced into a photodiode to convert the light intensity to a voltage. The CPT resonance spectrum can be generated by scanning the modulation frequency. In this setup, the CPT resonance signal can be evaluated against controllable parameters such as the intensity and frequency of the laser, polarization of the excitation beam, and the applied magnetic field at the Cs-vapor cell.

## VI. RESULTS AND DISCUSSIONS

### A. CPT spectra excited by $\sigma-\sigma$ and lin ∥ lin polarizations

Figure 3 shows typical CPT spectra excited by (a) $\sigma^--\sigma^-$ and (b) lin ∥ lin polarizations. The wavelength of the laser with a modulation frequency $f_m$ of near $f_{HFS}/2$ is tuned so that its optical frequencies of the two first-order sidebands correspond to the $|g\rangle$ to $|F'=3\rangle$ and the $|e\rangle$ to $|F'=3\rangle$ transitions, respectively. The sum of light intensities for the two sidebands is $5.1\ \mu W/mm^2$, and the magnetic field of $40\ \mu T$ is applied to resolve the degeneracy of the Zeeman sublevels. By scanning the modulation frequency in a width of $\pm 1.5\ MHz$, the transmitted light passed through the cell is detected as a function of $2f_m - f_{HFS}$, which we call the detuning frequency.

In Fig. 3 (a), there are six peaks that correspond to the CPT resonances for the $(m, m)$ transition except for $m=-3$, as shown in Table I, when Cs atoms are excited by $\sigma^--\sigma^-$ polarization. In Fig. 3 (b), seven symmetrical peaks derived from excitation by lin ∥ lin polarization are observed. Each CPT resonance excited by lin ∥ lin polarization is assigned to $(m-1, m+1)$ and $(m+1, m-1)$ transitions by Table I. The two CPT resonances overlap to be a single peak. Except for the central largest peak of the spectrum, other six peaks have wider linewidths, which arise from the fluctuations and inhomogeneity of the magnetic field. The central peaks caused by the (0, 0) or (-1, 1) and (1, -1) transitions have the largest amplitude and narrowest linewidth because the magnetic-field sensitivity is more than 100 times smaller than that of others.

## B. Spectra of central CPT resonance

Next, we investigate more precisely the absorption profile and the central peaks for CPT resonances in $\sigma - \sigma$ and lin ∥ lin excitation as a function of laser wavelength and the detuning frequency between sidebands. Our experiment was conducted at the sum of the sideband intensity of 0.43 μW/mm$^2$ and the magic magnetic field of 139 μT.

Figure 4 (a) and (b) indicate the absorption spectra in the frame of the wavelength of laser and the modulation frequency for the excitation of $\sigma - \sigma$ and lin ∥ lin polarizations, respectively. The detuning frequency $2f_m - f_{HFS}$ is swept in a range of 6.8 GHz with a step of 2.5 MHz. We can find two strong absorptions, when the wavelength of the laser modulated by $f_{HFS}/2$ is tuned to 894.593 nm or 894.590 nm, where atoms are excited to $F' = 3$ or $F' = 4$ from the ground hyperfine states, respectively. The other two absorptions correspond to the points where the two transitions such as a transition from $|F = 3\rangle$ to $|F' = 3\rangle$ and a transition from $|F = 4\rangle$ to $|F' = 4\rangle$ occur at the same time. These two absorptions do not achieve the two-photon Λ scheme.

The amplitude of the absorption with a detuning frequency of zero is shown in Fig. 4 (c). The absorption to the $F' = 4$ state is stronger than that of the $F' = 3$ state for both excitations. The half-width at a half-maximum (HWHM) of absorption line is 240 MHz for $\sigma - \sigma$ and 200 MHz for lin ∥ lin polarization. Thus, the width of the absorption line for the cell with a buffer gas pressure of 0.13 kPa is sufficiently narrower than the hyperfine splitting of the excited state, 1168 MHz.

CPT resonances cannot be detected in the above measurements because the linewidths for CPT resonances are generally much narrower than the step width for $2f_m$. When scanning $f_m$ in a narrow step, such as several tens Hz, CPT resonance for the $(m_g, m_e)$ transition can be detected at $2f_m = \Delta f_{m_g, m_e}$, if the dark state of $(m_g, m_e)$ exists under the experimental conditions applied. The transmission spectra in Fig. 4 (d) for $\sigma - \sigma$ and (e) for lin ∥ lin excitations are plotted against the detuning frequency, which is scanned with a span of 10 kHz and a step of 7 Hz. The bright points show the CPT resonances for the excitation to the $F' = 3$ or $F' = 4$. On each excitation, a single point is observed for $\sigma - \sigma$ excitation, but a number of points separated by 3.1 kHz are observed for lin ∥ lin excitation. As is explained in Sec. II, the CPT resonance for $\sigma - \sigma$ excitation is assigned to the (0, 0) transition and a pair of peaks observed for lin ∥ lin excitation is assigned to the (-1, 1) and (1, -1) transitions. Figure 4 (f) shows the CPT amplitudes as a function of wavelength of the laser. The amplitude of (0, 0) spectra at $F' = 4$ is larger than that at $F' = 3$. In contrast, (-1, 1) and (1, -1) spectra at $F' = 3$ are larger than those at $F' = 4$, which is consistent with those reported in [20].

Figure 4 (g) and (h) show the spectra for the CPT resonance to $F' = 3$ as a function of detuning frequency. The HWHM is about 0.20 kHz for the (0, 0) spectrum and 0.24 kHz for both (-1, 1) and (1, -1) spectra.

### C. CPT resonance in magnetic field

As stated in Sec. II, the frequencies of (0, 0), (-1, 1), and (1, -1) resonances vary slightly as the strength of the longitudinal magnetic field increases. Figure 5 (a) and (b) show shifts of CPT resonances for $\sigma - \sigma$ and lin ∥ lin excitations at several strengths of magnetic fields, from 8 to 470 µT. The intensity of excitation lights is 0.44 µW/mm$^2$. The CPT resonance for $\sigma - \sigma$ excitation is a single resonance. For lin ∥ lin excitation, however, one resonance at $B = 8$ µT splits into two components that separate as the strength of the magnetic field increases. The amplitudes of the CPT resonances are almost constant regardless of the strength of magnetic fields, except for at 8 µT, where two resonances are overlapped for lin ∥ lin excitation. The linewidths of each CPT resonances are 0.24 kHz in HWHM and clearly split at a magnetic field of more than 20 µT.

The behaviors of frequency shifts of the CPT resonances relative to the magnetic field are compared with Eqs. (1–4) of resonance frequencies for (0, 0), (-1, 1), and (1, -1) transitions. After that, the single CPT resonance excited by $\sigma - \sigma$ polarization is coincident to the (0, 0) transition, and a pair of CPT resonances excited by lin ∥ lin polarization is coincident to the (-1, 1) and (1, -1) transitions. Therefore, we can confirm that the CPT resonance of the (0, 0) transition does not occur for excitation by lin ∥ lin polarization, as stated in Sec. II. The frequency curves for (0, 0), (-1, 1), and (1, -1) transitions are given by solid lines in Fig. 6, together with those experimental values by closed circles. The calculated values are plotted with an offset frequency, which corresponds to shifts due to causes other than a magnetic field, such as the buffer gas shift of N$_2$ [27,28]. The experimental values are in good agreement with the calculated ones within the averaged residual frequency difference of 27 Hz. Focusing on (-1, 1) resonance, the frequency shift is determined to be

$$\Delta f_{-1,1}(B) / \text{kHz} = f_{\text{HFS}} / \text{kHz} + 1.42(2) + 40.9(3) \times 10^{-6} \left( B / \mu\text{T} - 141(2) \right)^2, \tag{10}$$

which confirms the magic magnetic field of 139 µT with a reasonable accuracy. The constant term includes the offset frequency.

Thus, (-1, 1) resonance frequency is exactly insensitive to the magnetic field at 139 µT. In a condition where the HWHM of the CPT spectrum is narrower than half of the separation between the two resonances, (-1, 1) resonance can be observed without overlapping with (1, -1) resonance.

Regarding applications of the CPT resonance to atomic clocks, a transition insensitive to the fluctuation of the longitudinal magnetic field is preferred. In miniaturized CPT clocks, the vapor cell is usually surrounded by a magnetic shield with a shielding factor of several tens [3], so the magnitude of the magnetic field fluctuation in the vapor cell is estimated to be about 1 µT. Conventional CPT clocks use the (0, 0) transition as the frequency reference at a typical bias magnetic field of about 25 µT [3] to resolve the degeneracy of the Zeeman sublevels so that the frequency sensitivity to magnetic fields is about 2.1 Hz/µT. In contrast, the CPT resonance frequency of the (-1, 1) transition varies only 0.04 Hz against the variation of $\pm 1$ µT if we measure it at the magic magnetic field.

### D. Dependence of CPT resonance on excitation intensity

(-1, 1) and (1, -1) resonances excited by lin ∥ lin polarization are measured at different excitation intensities up to 4.4 µW/mm$^2$, as shown in Fig. 7. Here, the laser wavelength is tuned to the transition from the ground state to the $F' = 3$ state. The interval between the (-1, 1) and (1, -1) resonances is 3.1 kHz at $B = 139$ µT. We confirmed that the light shift observed in these measurements is less than 0.5 kHz for both resonances. To analyze the CPT resonances, the sum of two Lorentzian curves with the same amplitude and the same width are fitted to the CPT spectra. The behavior of the (0, 0) CPT resonance excited to the $F' = 3$ state by $\sigma - \sigma$ polarization was also investigated for the comparison.

The HWHM of (-1, 1) and (0, 0) resonances are shown in Fig. 8 (a). In both excitations, monotonic broadenings against the excitation intensity start from about 0.15 kHz at 0.1 µW/mm$^2$. The gradients of (-1, 1) and (0, 0) resonance are 0.2 kHz/(µW/mm$^2$) and about 0.1 kHz/(µW/mm$^2$), respectively. As a result, the width of (-1, 1) resonance becomes about 1.7 times larger than that of (0, 0) resonance at 4 µW/mm$^2$. Vanier derived a relation wherein the initial broadening is equal to the dephasing parameter $\gamma_c$, and the gradient is the square of the Rabi frequency (which is proportional to the intensity) divided by optical coherence [8].

Figure 8 (b) and (c) show the absorption of Cs and amplitudes of the CPT resonance against the excitation intensity. In $\sigma - \sigma$ excitation, both absorption and CPT amplitude moderately saturate at high intensity. In contrast, the CPT amplitude in lin ∥ lin excitation is almost proportional to the intensity. Thus, (-1, 1) resonances are detectable in higher amplitude at higher intensity.

The behavior of the CPT resonance in the three-level system $|g\rangle$, $|e\rangle$, and $|i\rangle$ is described using the density matrix following the Liouville equation. The calculation based on the three-level model shows the linear dependence of the CPT amplitude on the excitation intensity asymptotically. However, atoms such as Cs include several energy levels besides the three levels; atoms excited to the $|i\rangle$ state under circular polarization decay into other levels and are trapped for a time of the order of $1/\gamma_c$, as shown in Fig. 8 (d). For $\sigma - \sigma$ excitation, the amplitude of the CPT resonance saturates when the trap state exists. Taichenachev *et al.* predicted that the (-1, 1) resonance excited by lin ∥ lin polarization of Cs atoms will saturate because of the trap state [17].

We calculate the steady-state solution of the density-matrix using the Liouville equation on the three-level system [8] for the (-1, 1) resonance excited by lin ∥ lin polarization and on the four-level system for the (0, 0) resonance excited by $\sigma - \sigma$ polarization. In the calculation, the following conditions are used: ground level decay rate $\gamma_p = 0.3$ kHz , dephasing rate $\gamma_c = 0.15$ kHz , population relaxation of the excited state $\Gamma = 240$ MHz for $\sigma - \sigma$, 200 MHz for lin ∥ lin, and optical coherence $\Gamma_c = \Gamma/2$. It is assumed that the equilibrium value of the ground-state levels is 1/2 for the three-level model and 1/3 for the four-level model. Results are shown by solid curves in Fig. 8 (a)–(c). The calculation curves are fitted to the experimental data in magnitude. We conclude that the CPT resonance excited by $\sigma - \sigma$ polarization is in good agreement with the curves obtained by the four-level model with a trap state, while CPT resonance excited by lin ∥ lin polarization is described well by the three-level model without a trap state. In other words, the CPT resonance excited by lin ∥ lin polarization to the $F' = 3$ state in $^{133}$Cs is not disturbed by any trap state, so the (-1, 1) resonance via $F' = 3$ is suitable for miniature atomic clocks.

In this paper, we used simple equations on Cs atoms to show that the behavior of the CPT resonance excited by lin ∥ lin polarization would be described by the three-level model without a trap state, while that excited by $\sigma - \sigma$ polarization would be described by the four-level model with a trap state. The theoretical calculation of the CPT resonance of $^{87}$Rb excited by different polarization schemes was developed by Warren *et al*. using the Liouville density-matrix equation taking into account all relevant Zeeman sublevels, and compared with the experimental data successfully [19]. We are now developing a similar calculation of the CPT resonance of $^{133}$Cs taking into account all 32 Zeeman sublevels of the $D_1$ line.

## V. CONCLUSION

In summary, we have experimentally demonstrated that the two (-1, 1) and (1, -1) CPT resonances are generated between the ground hyperfine levels on the $D_1$ line of $^{133}$Cs using a two-photon $\Lambda$ scheme excited by lin ∥ lin polarizations. Using a Cs-vapor cell with a 0.13 kPa of $N_2$ buffer gas at room temperature, the width (HWHM) of the CPT resonance is about 0.24 kHz with an excitation intensity of $0.4$ μW/mm$^2$, so that the two resonances are clearly separated from each other under the longitudinal magnetic field more than 20 μT . We also have confirmed that (0, 0) CPT resonance is not induced by lin ∥ lin excitation, because the common dark state for two $\Lambda$ schemes of (0, 0) CPT resonance does not exist [12].

The (-1, 1) and (1, -1) CPT resonances, as well as (0, 0) CPT resonance, are magnetic field-insensitive transition, because their first-order Zeeman shifts are nearly cancelled. Their frequency dependences on the longitudinal magnetic field were measured precisely and shown to be in good agreement with the Breit-Rabi formula [13] within 27 Hz. Notably, the frequency sensitivity of (-1, 1) CPT resonance on the variation of the magnetic field becomes minimum at a "magic" magnetic field of 139 μT. The frequency sensitivity on the magnetic field is 0.04 Hz/μT, which is 50 times lower than that of (0, 0) CPT resonance.

The amplitude of the (-1, 1) CPT resonance excited to $F'=3$ state by lin ∥ lin polarizations was investigated as a function of the excitation intensity, together with the (0, 0) CPT resonance excited to $F'=3$ state by $\sigma-\sigma$ polarizations for comparison. The fact that the amplitude of (0, 0) CPT resonance excited by $\sigma-\sigma$ polarization is saturated as the excitation intensity increases would be explained well by the four-level model with a trap state [8]. On the other hand, the amplitude of (-1, 1) CPT resonance excited by lin ∥ lin polarization increases without saturation as the excitation intensity increases. The fact would be explained by the three-level model without a trap. The amplitude of (-1, 1) CPT resonance becomes higher than that of (0, 0) CPT at the excitation intensity higher than 2.5 μW/mm$^2$.

Thus, (-1, 1) CPT resonance on the $D_1$ line of Cs atom excited by a simple lin ∥ lin scheme will be one of the best candidates for realizing the miniature atomic clock based on CPT phenomenon.

## ACKNOWLEDGMENTS

This study was supported by Innovative Science and Technology Initiative for Security Grant Number JPJ004596, ATLA, Japan. A.M. thanks Dr. Katsumi Irokawa of Tokyo university of Science, for his kind guidance on the calculation.

TABLE I. $\Lambda$ schemes for the CPT resonance on the $D_1$ line of $^{133}\text{Cs}$, together with the polarization scheme. $m_g$ and $m_e$ are magnetic quantum numbers of the lower and the upper levels in the ground hyperfine state, $m_{F'}$ and $F'$ are the magnetic quantum number and total angular momentum of the excited state, respectively, and $m$ is the allowed magnetic quantum number. Dark state conditions for $m=0$ for the $\Lambda$ schemes are given.

| CPT resonance $(m_g, m_e)$ | Polarization of $(\omega_1, \omega_2)$ | $m_{F'}$ | $F'$ | $m$ | Dark state condition at $m=0$ and $\varepsilon_1 = \varepsilon_2$ |
|---|---|---|---|---|---|
| $(m, m)$ | $(\sigma^+, \sigma^+)$ | $m+1$ | 3 | -3, -2, …, 2 | $e^{i\theta}$ |
|  |  |  | 4 | -3, -2, …, 3 | $e^{i\theta}$ |
|  | $(\sigma^-, \sigma^-)$ | $m-1$ | 3 | -2, -1, …, 3 | $-e^{-i\theta}$ |
|  |  |  | 4 | -3, -2, …, 3 | $-e^{-i\theta}$ |
| $(m-1, m+1)$ | $(\sigma^+, \sigma^-)$ | $m$ | 3 | -2, -1, …, 3 | $-e^{-i\theta}\sqrt{3/5}$ |
|  |  |  | 4 | -2, -1, …, 3 | $e^{-i\theta}\sqrt{3/5}$ |
| $(m+1, m-1)$ | $(\sigma^-, \sigma^+)$ | $m$ | 3 | -3, -2, …, 2 | $e^{i\theta}\sqrt{3/5}$ |
|  |  |  | 4 | -3, -2, …, 2 | $-e^{i\theta}\sqrt{3/5}$ |

Figure Captions

FIG. 1 (Color online) Hyperfine structures on $D_1$ line of $^{133}$Cs, together with two-photon $\Lambda$ scheme transitions between $|g\rangle$ and $|e\rangle$ states via $|i\rangle$ state. $\Lambda$ schemes are displayed with dashed and solid lines which are excited by $\sigma-\sigma$ and lin ∥ lin polarizations, respectively.

FIG. 2 (Color online) Experimental setup for investigation of the CPT resonance excited by $\sigma-\sigma$ or lin ∥ lin polarizations. VOA: Variable optical attenuator, HWP: Half-wave plate, PBS: Polarizing beam splitter, QWP: Quarter-wave plate, and PD: Photodetector. Cell contains $^{133}$Cs and 0.13 kPa of $N_2$ buffer gas.

FIG. 3 (Color online) CPT spectra excited to $F'=3$ state by (a) $\sigma-\sigma$ and (b) lin ∥ lin polarizations as a function of detuning frequency. Both are measured with the excitation intensity of 5.1 μW/mm$^2$ at magnetic field of 40 μT.

FIG. 4 (Color online) Absorption map of $^{133}$Cs atoms for the incident lights with (a) $\sigma-\sigma$ and (b) lin ∥ lin polarizations in the plane of the laser wavelength and the coarse detuning frequency between sidebands. Color scale shows the intensity of the transmitted light. (c) Amplitude of absorptions as a function of the carrier wavelength. CPT map excited by (d) $\sigma-\sigma$ and (e) lin ∥ lin polarizations in the plane of the laser wavelength and the fine detuning frequency between sidebands. (f) Amplitudes of CPT resonance as a function of the carrier wavelength. (g) Spectrum of (0, 0) CPT resonance excited to $F'=3$. (h) Spectra of (-1, 1) and (1, -1) CPT resonances excited to $F'=3$. Spectra are measured with an excitation intensity of 0.43 μW/mm$^2$ at magnetic field of 139 μT.

FIG. 5 (Color online) CPT resonance spectra depending on the strength of the longitudinal magnetic field. (a) (0, 0) CPT resonance excited by $\sigma-\sigma$ polarization. (b) (-1, 1) and (1, -1) CPT resonances excited by lin ∥ lin polarization. Spectra are measured with an excitation intensity of 0.44 μW/mm$^2$.

FIG. 6 (Color online) Resonance frequencies of (1, -1), (0, 0), and (-1, 1) CPT spectra depending on the strength of a longitudinal magnetic field. Closed circles represent experimental values. Black solid lines are the calculated transition frequencies based on the Breit-Rabi formula, which includes a specific offset of 2.2 kHz.

FIG. 7 (Color online) CPT spectra excited by lin ‖ lin polarization to F' = 3 state measured at $B = 139\ \mu T$. In sequence from top, excitation intensity is 4.4, 2.9 1.8, 1.1, 0.6, 0.3, 0.2 $\mu W/mm^2$, respectively.

FIG. 8 (Color online) Comparison of the experimental data with the calculated curves for the CPT resonances excited by $\sigma - \sigma$ (circle) and lin ‖ lin (square) polarization. (a) Linewidth (half width at half maximum) of CPT resonance versus intensity, (b) Absorption of the incident light, and (c) Amplitudes of CPT signals. The wavelength of the incident light is tuned to the excitation to $F' = 3$ state. A magnetic field of 139 $\mu T$ is applied. (d) Four-level models taking into account the trap state is used in the analysis for the $\sigma - \sigma$ excitation while three-level models where the trap state $|b\rangle$ is omitted are applied for the lin ‖ lin excitation.

FIG. 1

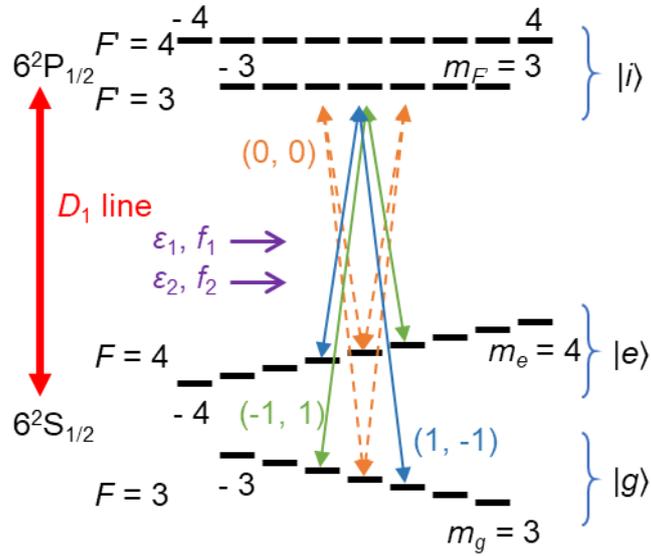

FIG. 2

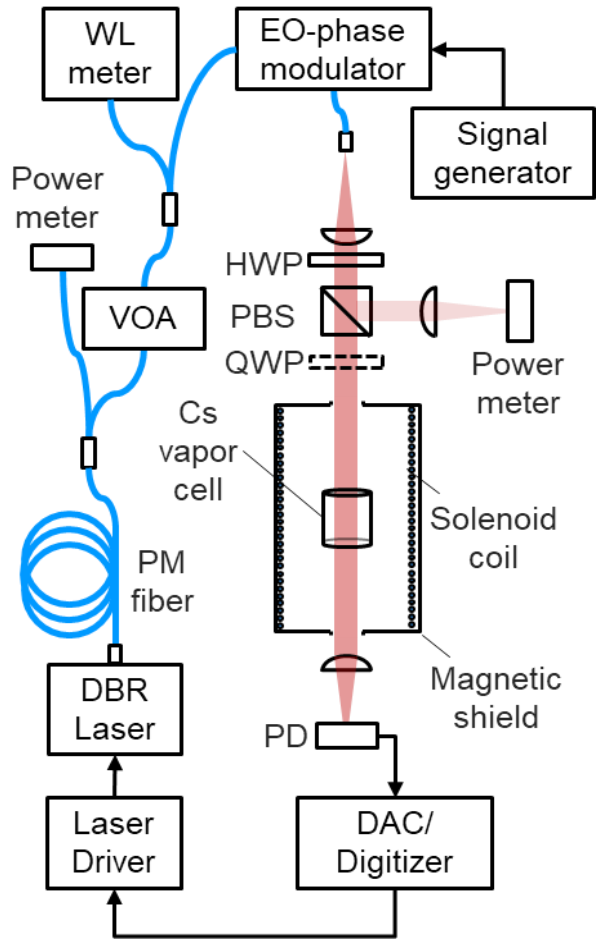

FIG. 3

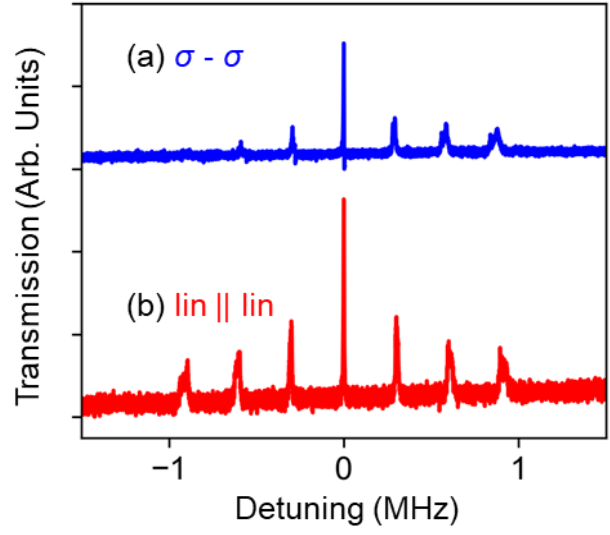

FIG.4

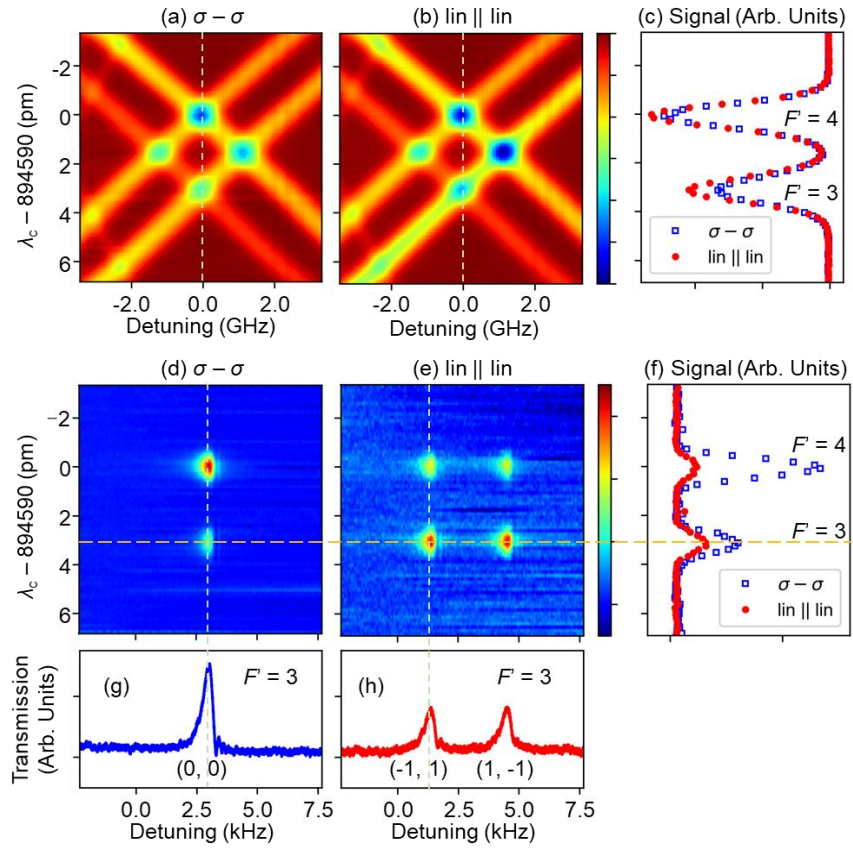

FIG.5

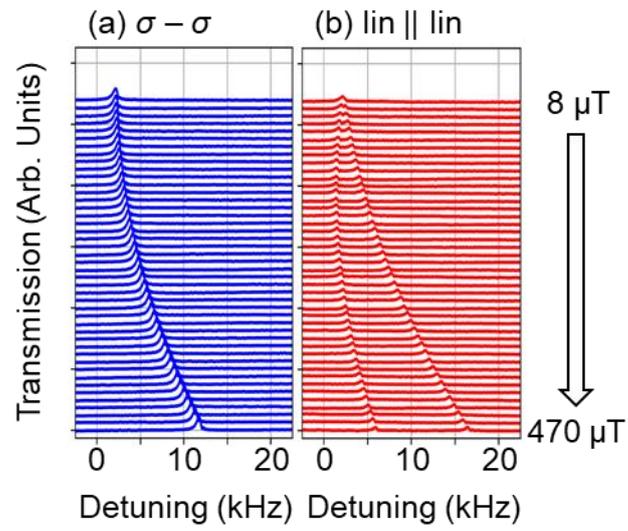

FIG.6

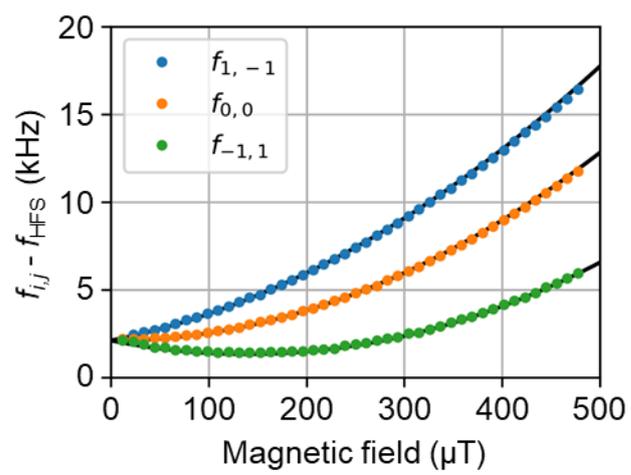

FIG.7

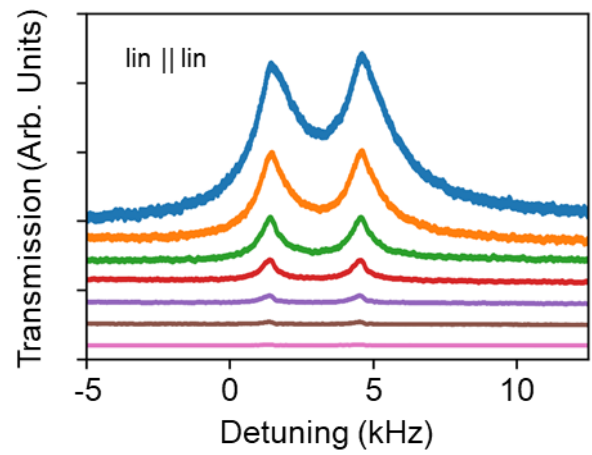

FIG. 8

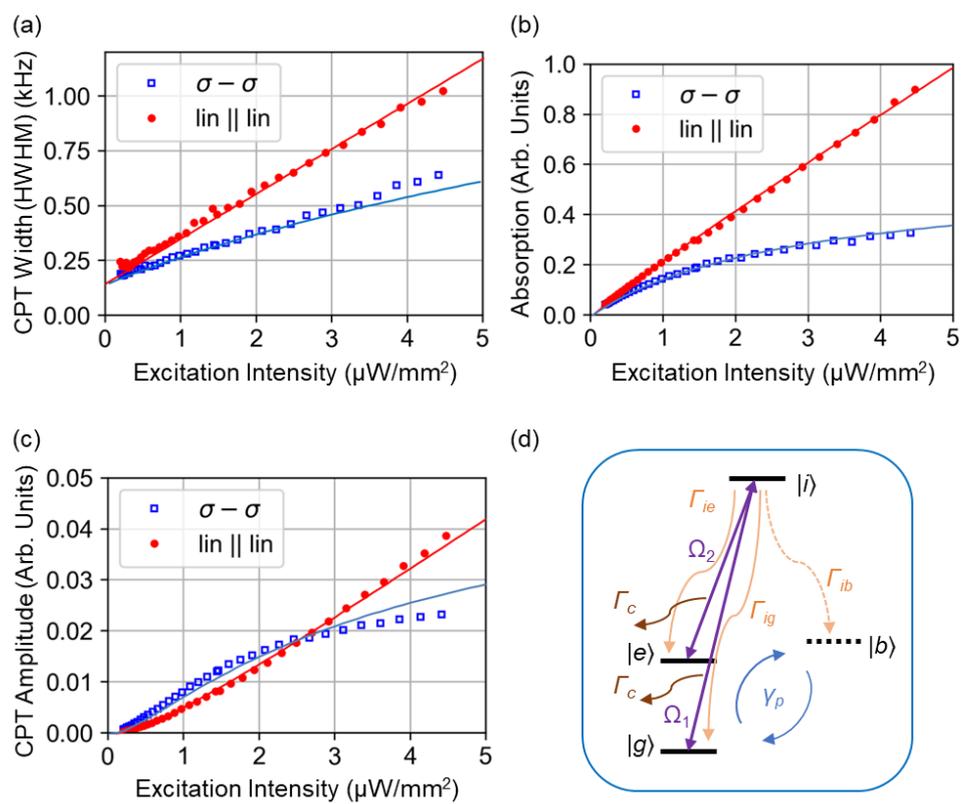